\documentclass[prl,showpacs,amsmath,amssymb, twocolumn, 10pt]{revtex4}

\usepackage{amsthm}
\usepackage{dcolumn}
\usepackage{bm}
\usepackage{graphicx}

\begin{document}

\newtheorem{corollary}{Corollary}
\newtheorem{definition}{Definition}
\newtheorem{example}{Example}
\newtheorem{lemma}{Lemma}
\newtheorem{proposition}{Proposition}
\newtheorem{theorem}{Theorem}
\newtheorem{fact}{Fact}
\newtheorem{property}{Property}
\newcommand{\bra}[1]{\langle #1|}
\newcommand{\ket}[1]{|#1\rangle}
\newcommand{\braket}[3]{\langle #1|#2|#3\rangle}
\newcommand{\ip}[2]{\langle #1|#2\rangle}
\newcommand{\op}[2]{|#1\rangle \langle #2|}

\newcommand{\tr}{{\rm tr}}
\newcommand {\E } {{\mathcal{E}}}
\newcommand {\F } {{\mathcal{F}}}
\newcommand {\diag } {{\rm diag}}

\title{Distinguishing Arbitrary Multipartite Basis Unambiguously Using Local Operations and Classical Communication}
\author{Runyao Duan}
\email{dry@tsinghua.edu.cn}
\author{Yuan Feng}
\email{feng-y@tsinghua.edu.cn}
\author{Zhengfeng Ji}
\email{jizhengfeng98@mails.tsinghua.edu.cn}
\author{Mingsheng Ying}
\email{yingmsh@tsinghua.edu.cn}

\affiliation{State Key Laboratory of Intelligent Technology and
Systems, Department of Computer Science and Technology, Tsinghua
University, Beijing, China, 100084}

\date{\today}

\begin{abstract}
We show that an arbitrary basis of a multipartite quantum state
space consisting of $K$ distant parties such that the $k$th party
has local dimension $d_k$ always contains at least $N=\sum_{k=1}^K
(d_k-1)+1$ members that are unambiguously distinguishable using
local operations and classical communication (LOCC). We further show
this lower bound is optimal by analytically constructing a special
product basis having only $N$ members unambiguously distinguishable
by LOCC. Interestingly, such a special product basis not only gives
a stronger form of the weird phenomenon ``nonlocality without
entanglement", but also implies the existence of locally
distinguishable entangled basis.
\end{abstract}


\maketitle

Suppose we are given a quantum system whose state is secretely
chosen from a finite set of pre-specified  unit vectors. Because of
the limitations of quantum mechanics, the state of the system can be
identified with certainty if and only if these vectors are mutually
orthogonal. Remarkably, when these vectors are nonorthogonal but
linearly independent, the state of the system can also be identified
unambiguously with some nonzero success probability \cite{Che98}.
Consequently, any orthogonal (or nonorthogonal) basis always can be
exactly (respectively, unambiguously) discriminated when there is no
restrictions on the quantum measurements one can perform.

However, the situation becomes very complicated when the given
quantum system is shared by a finite number of distant parties,
where each party holds a piece of the whole system and can perform
local operations and classical communication (LOCC) only
\cite{PW91}. The number of states one can locally discriminate
significantly decreases. Perhaps the most surprising discovery in
this field is due to Bennett and collaborators in Ref.
\cite{BDF+99}, where they exhibited nine $3\otimes 3$ orthogonal
product states that are not exactly distinguishable using LOCC, thus
initiated the study of the famous phenomenon ``nonlocality without
entanglement". Another fundamental contribution was made by Walgate
\textit{et al.} in Ref. \cite{WSHV00}, where it was demonstrated
that any two multipartite orthogonal pure states, whether entangled
or not, can always be exactly discriminated using LOCC. Since then,
many interesting works on the local distinguishability of quantum
states have been done and two kinds of results have been reported
\cite{BDM+99, DMS+00,TDL01,GKR+01,VSPM01,
WH02,CL04,HSSH03,CHE04,WAT05,HMM+06, BW06}. The first kind of
results follows Ref. \cite{BDF+99} to show that certain set of
quantum states cannot be discriminated by LOCC. The second kind of
results is, similar to Ref. \cite{WSHV00}, to identify sets of
states that are locally distinguishable. In particular, an
orthonormal basis that is unambiguously distinguishable by LOCC must
be a product basis, as shown by Horodecki \textit{et al.}
\cite{HSSH03}. Another very interesting result is recently obtained
by Bandyopadhyay and Walgate in Ref. \cite{BW06}, where they showed
that among any three linearly independent pure states, there always
exists one state that can be unambiguously determined using LOCC.

In this Letter we consider the local distinguishability of a general
nonorthogonal multipartite basis. Our main result is that arbitrary
multipartite basis has at least $N$ members that are unambiguously
distinguishable by LOCC. This lower bound is shown to be tight by
explicitly constructing a product basis which has only $N$ members
that are unambiguously distinguishable by LOCC (Theorem $2$). The
significance of our result is that it gives a universal tight lower
bound on the number of the locally unambiguously distinguishable
members in arbitrary multipartite basis and thus provides new
insight into the local distinguishability of a multipartite basis.
We further obtain an equivalence between locally distinguishable
entangled basis (DEB) (basis having entangled states as members) and
indistinguishable product basis (IPB) (basis having product states
as members) (Theorem $3$), and present explicit constructions of
such special basis. In particular, the existence of locally
indistinguishable product basis gives a stronger form of
``nonlocality without entanglement". Furthermore, these results
yield the following counterintuitive conclusion about the relations
among orthogonality, entanglement, and local distinguishability:
Orthogonal states are not always more easily locally distinguishable
than nonorthogonal ones, sometimes less orthogonality and more
entanglement (or vice versa) may enhance the local
distinguishability of a set of quantum states. Most notably, in
proving our main results, we introduce a useful notion named
unextendible bases and employ it as a tool to study the local
distinguishability of quantum states. As a byproduct, we obtain an
interesting connection between the local unambiguous
distinguishability and unextendible bases (Theorem $1$).

Let us now begin to introduce the notion of unextendible bases,
which is a generalization of the orthogonal unextendible product
bases that was first introduced by Bennett \textit{et al.}
\cite{BDM+99} and was then extensively studied in Ref.
\cite{DMS+00}. The key difference is that in Refs.
\cite{BDM+99,DMS+00} only orthogonal product states are considered
while here arbitrary states (entangled or unentangled) may be
involved.  Let $\mathcal{H}=\otimes_{k=1}^K\mathcal{H}_{k}$ be a
multipartite quantum system with $K$ parties. Each $\mathcal{H}_k$
is a $d_k$-dimensional Hilbert space. Sometimes we use the notation
$d_1\otimes \cdots \otimes d_K$ for $\mathcal{H}$. Let $S$ be a
subset of $\mathcal{H}$. Then $span(S)$ is the subspace spanned by
the vectors in $S$, and $S^\perp$ represents the orthogonal
complement of $S$.

{Definition 1 (unextendible bases):} Let
$S=\{\Psi_1,\cdots,\Psi_m\}$ be a collection of $m$ linearly
independent quantum states on $\mathcal{H}$. $S$ is said to be an
\textit{unextendible bases} (UB) if $S^\perp$ contains no product
state; otherwise $S$ is said to be extendible. Furthermore, $S$ is
said to be a \textit{genuinely unextendible bases}(GUB) if it is
unextendible and any proper subset of $S$ is extendible. In
particular, when $S$ is a collection of product states, we use the
notions unextendible product bases (UPB) and genuinely unextendible
product bases (GUPB) instead of unextendible bases and genuinely
unextendible bases, respectively.

It follows directly from the above definition that any orthogonal
UPB is necessarily a GUPB. However, when nonorthogonal states are
considered, there does exist UPB that is not a GUPB. See Example $2$
below for an explicit instance.

A notion closely related to unextendbile bases is the
\textit{completely entangled subspace} introduced by Parthasarthy
\cite{PAR04}: $S$ is a UB implies $S^\perp$ is completely entangled;
conversely, $S$ is completely entangled indicates that any basis of
$S^\perp$ constitutes a UB. Such a correspondence suggests that some
works that have been done for completely entangled subspaces may be
useful in studying UB.

The following lemma provides a lower bound on the size of a UB.

{Lemma 1:} Any UB on $\mathcal{H}$ must have at least
$N=\sum_{k=1}^K (d_k-1)+1$ members.

{Proof:} Let $S$ be a UB on $\mathcal{H}$. Then the cardinality of
$S$ is just the dimension of $span(S)$. For simplicity, we write
directly $dim(S)$ for the cardinality of $S$. Then
$dim(S)+dim(S^\perp)=dim(\mathcal{H}).$ By the definition, $S^\perp$
is a completely entangled subspace of $\mathcal{H}$. We notice that
it has been proven by Parthasarthy \cite{PAR04} that the maximal
dimension of a completely entangled subspace of $\mathcal{H}$ is
$dim(\mathcal{H})-N$. Applying this result we have that $dim(S)=
dim(\mathcal{H})-dim(S^\perp)\geq N.$ That completes the proof.
\hfill $\square$

It is well known that the construction of orthogonal UPB is an
extremely difficult task \cite{BDM+99,DMS+00}. Interestingly,
constructing nonorthogonal UB and UPB is very simple. We shall need
the following counting lemma which is essentially due to Bennett
\textit{et al.} \cite{BDM+99}:

{Lemma 2:} Let
$S=\{\ket{\psi_j}=\otimes_{k=1}^K\ket{\varphi_{kj}}:1\leq j\leq N\}$
be a collection of product states on $\mathcal{H}$. If for each
$1\leq k\leq K$, any subset of $\{\ket{\varphi_{kj}}:1\leq j\leq
N\}$ with $d_k$ members spans $\mathcal{H}_k$, then $S$ is a GUB for
$\mathcal{H}$.

Assisting with Lemma 2,  we are now in a position to present an
explicit construction of GUPB with minimal cardinality.

{Example 1 (minimal UPB):~~~}For each $1\leq k\leq K$, let
$\{\ket{j}:0\leq j\leq d_k-1\}$ be an orthonormal basis for
$\mathcal{H}_k$. For each $1\leq k\leq K$ and $x\in \mathcal{C}$,
define
$\ket{\varphi_{k}(x)}=N_k(x)^{-1}\sum_{j=0}^{d_k-1}x^j\ket{j}$,
where $N_k(x)=\sqrt{\sum_{j=0}^{d_k-1}|x|^{2j}}$ is the normalized
factor. For simplicity, let
$\ket{\varphi_k(\infty)}=\lim_{x\rightarrow
\infty}\ket{\varphi_k(x)}=\ket{d_k-1}$, and let
$\mathcal{C}_{+}=\mathcal{C}\cup\{\infty\}$. A key property of
$\ket{\varphi_k(x)}$ is that for any pairwise different elements
$x_0,\cdots, x_{d_k-1}$ from $\mathcal{C}_{+}$,
$\ket{\varphi_k(x_0)},\cdots, \ket{\varphi_k(x_{d_k-1})}$ are
linearly independent and thus form a basis for $\mathcal{H}_k$. This
can be seen from the nonsingularity of Van der Monde matrix
$[x_m^n]$, $0\leq m,n\leq d_{k}-1$. (Note that the nonsingularity is
also valid when one of $x_m$ is $\infty$).

For $\lambda=(\lambda_1,\cdots, \lambda_K)\in \mathcal{C}_+^{K}$, we
define $\ket{\psi(\lambda)}=\ket{\varphi_1(\lambda_1)}\otimes\cdots
\otimes \ket{\varphi_K(\lambda_K)}.$ Take an index set
$I=\{\lambda^{(1)},\cdots, \lambda^{(N)}\}\subseteq\mathcal{C}_+^K$
such  that any two elements in $I$ are entrywise distinct, i.e.,
$\lambda_{k}^{(m)}\neq \lambda_{k}^{(n)}$ for any $1\leq m<n\leq N$
and $1\leq k\leq K$. Then the set $\{\ket{\psi(\lambda)}:\lambda\in
I\}$ is a UPB on $\mathcal{H}$.

By Lemma 2, we only need to show that for each $1\leq k\leq K$, any
$d_k$ members of $\{\ket{\varphi_k(\lambda^{(j)}_k)}:1\leq j\leq
N\}$ are linearly independent and thus form a basis for
$\mathcal{H}_k$. This clearly holds by the pairwise distinctness of
$\{\lambda_k^{(j)}:1\leq j\leq N\}$ and by the special form of
$\ket{\varphi_k(.)}$.

In particular, take $K=2$, $d_1=d_2=2$,
$I=\{(0,0),(1,1),(\infty,\infty)\}$, then
$\{\ket{0}\ket{0},\ket{+}\ket{+},\ket{1}\ket{1}\}$ is a minimal UPB
on $2\otimes 2$. \hfill $\square$

A special case of the above construction ($\lambda$ satisfies
$\lambda_1=\cdots=\lambda_K$) was first given by Parthasarthy and
then was considerably studied by Bhat \cite{PAR04}. Unfortunately,
their method can only yield $N$ linearly independent vectors. The
above construction is much more general and can be used to construct
a product basis for $\mathcal{H}$, as we will see latter.

We shall present a connection between local unambiguous
distinguishability and unextendible bases. The following lemma,
which is a simplified version of a more general result due to
Chefles \cite{CHE04}, indicates that the condition for unambiguous
discrimination is much more complicated when only LOCC operations
are allowed. Recently an alternative proof was obtained in Ref.
\cite{BW06}.

 {Lemma 3:} Let $S=\{\Psi_1,\cdots, \Psi_m\}$
be a collection of $m$ quantum states on $\mathcal{H}$. Then $S$ can
be unambiguously discriminated by LOCC  if and only if for each
$1\leq k\leq m$, there exists a product detecting state
$\ket{\psi_k}$ such that $\ip{\Psi_j}{\psi_k}=0$ for $j\neq k$ and
$\ip{\Psi_k}{\psi_k}\neq 0$.

It should be noted that the proof of Lemma $3$ is not constructive.
So far there is no feasible way to determine the existence of the
product detecting states \cite{CHE04, BW06}. Consequently, for a
given set of quantum states, it is highly nontrivial to determine
the local distinguishability of these states.

Now we are ready to investigate the local distinguishability of a
UB. It has been shown by Bennett \textit{et al.} \cite{BDM+99} that
the members of an orthogonal UPB cannot be exactly discriminated by
LOCC. But clearly they are unambiguously distinguishable by LOCC, as
we can always choose the state itself as the corresponding detecting
state. Interestingly, this property holds for any GUB.

{Theorem 1:} A UB can be unambiguously discriminated by LOCC if and
only if it is a GUB.

{Proof:} Let $S=\{\Psi_1,\cdots,\Psi_m\}$ be a GUB. Consider the set
$S_k=S-\{\Psi_k\}$. $S_k$ is a proper subset of $S$ and thus is
extendible. So there exists a product state $\ket{\psi_k}$ such that
$\ip{\psi_k}{\Psi_j}=0$ for any $j\neq k$. We claim that
$\ip{\psi_k}{\Psi_k}\neq 0$. Otherwise, it holds $\ket{\psi_k}\in
S^\perp$. This contradicts the assumption that $S$ is unextendible.
Hence $\ket{\psi_k}$ is exactly a product detecting state for
$\Psi_k$. That proves the unambiguous distinguishability of $S$.

Now assume $S$ is a UB but not a GUB. So there is some proper subset
$S'$ of $S$ that is unextendible. In other words, $S'\subset S$ is
also a UB. Take $\Psi\in S-S'$. It is easy to see there cannot be a
product state $\ket{\psi}$ that is orthogonal to the vectors in
$S'$. Thus $\Psi$ cannot have a product state as its detector, and
cannot be identified with a nonzero probability from $S-\{\Psi\}$.
\hfill $\square$

Let us check some interesting consequences of Theorem 1. Suppose $S$
is a subspace spanned by a GUB with dimension $N$. Then any basis
for $S$ is also a GUB. Thus it follows from Theorem 1 that any basis
for $S$ is unambiguously distinguishable by LOCC. Interestingly,
recently Watrous found another kind of special bipartite subspace
having no basis exactly distinguishable by LOCC \cite{WAT05}.

Suppose now that $S$ is a UB but not a GUB. By the above theorem,
$S$ is not unambiguously distinguishable by LOCC. On the other hand,
$S$ always contains a proper subset, say $S'$, that is a GUB. Thus
$S'$ can be unambiguously discriminated by LOCC. An important
special case is when $S$ is a  basis for $\mathcal{H}$. Combining
Theorem 1 with Lemma 1, we arrive at the following central result:

{Theorem 2:} Any basis for $\mathcal{H}$ has at least $N$ members
unambiguously distinguishable by LOCC. Furthermore, the lower bound
$N$ is tight in the sense there exists  a basis for which any $N+1$
members cannot be unambiguously discriminated by LOCC.

Remark: In the case when $K=1$ (unipartite), $N$ is reduced to $d_1$
and we recover the well-known result that any basis can be
unambiguously discriminated by unconstrained measurements
\cite{Che98}.

Proof: Notice that any basis for $\mathcal{H}$ is a UB and thus
contains a GUB. By Lemma 1 and Theorem 1, such a GUB has a size at
least $N$ and is unambiguously distinguishable by LOCC. That
completes the proof of the first part. Now we consider the second
part. We shall give an explicit construction of the basis that has
only $N$ members unambiguously distinguishable by LOCC. For $x\in
\mathcal{C}$, define
$$\Psi(x)=\ket{\psi(x^{d_2\cdots d_K},\cdots, x^{d_K},x)},$$
where $\ket{\psi(.)}$ is the same as in Example 1. After some
algebraic manipulations we have
$$\Psi(x)=N(x)^{-1}\sum_{j=0}^{d-1}x^j\ket{j}_{\mathcal{H}},$$
where $d=d_1\cdots d_K$, $j=\sum_{k=1}^K j_k(d_{k+1}\cdots d_K)$,
$\ket{j}_{\mathcal{H}}=\otimes_{k=1}^K\ket{j_k}_{\mathcal{H}_k}$,
$0\leq j_k\leq d_{k}-1$, and $N(x)$ is a normalized factor. On the
one hand, take $d$ pairwise distinct complex numbers $x_1,\cdots,
x_d$ such that
$$x_m^{\prod_{j=k+1}^K d_{j}}\neq x_n^{\prod_{j=k+1}^K
d_{j}}, ~1\leq m<n\leq d, ~1\leq k\leq K.$$ Then by the
nonsingularity of Von der Monde matrix, we have that
$S=\{\Psi(x_1),\cdots,\Psi(x_d)\}$ is a  basis for $\mathcal{H}$. On
the other hand, by Example 1, any subset of $S$ with $N$ members
constitutes a GUB for $\mathcal{H}$. So any subset $S'$ of $S$ with
$N+1$ members is a UB but not a GUB. The indistinguishability of
$S'$ follows from Theorem 1.\hfill $\square$

One may naturally expect a stronger form of Theorem 2: Any $N$
linearly independent states of $\mathcal{H}$ can be unambiguously
discriminated by LOCC. Unfortunately, this cannot hold even for
$2\otimes 2$ states, as we have the following result: For any
$\mathcal{H}=\otimes_{k=1}^K \mathcal{H}_k$, there always exist
three \textit{orthogonal} pure states $\Psi_1,\Psi_2$, and $\Psi_3$
that are not unambiguously distinguishable by LOCC. A simple
instance is as follows (see Ref. \cite{BW06} for a similar
construction):
\begin{equation}\label{threestates}
\Psi_1=\ket{x},~\Psi_{2,3}=\frac{\ket{0}^{\otimes
K}\pm\ket{1}^{\otimes K}}{\sqrt{2}},
\end{equation}
where $x$ is a  $K$-bit string such that $x\neq 0^K,1^K$. We can
easily verify that $\Psi_{2(3)}$ cannot have a product detecting
state. Thus $\Psi_1$, $\Psi_2$, and $\Psi_3$ are not unambiguously
distinguishable by LOCC. This is an example of $K$ qubits. Obviously
it can also be treated as an example on any composite quantum system
consisting of $K$ parties.

Notice that any $d+1$ orthogonal maximally entangled states on
$d\otimes d$ cannot be exactly distinguishable by LOCC
\cite{GKR+01}. Interestingly, by Theorem 2 we conclude immediately
at least $2d-1$ maximally entangled states can be unambiguously
discriminated by LOCC. An explicit construction is as follows. Let
$\{\Phi_{m,n}:0\leq m,n\leq d-1\}$ be the canonical maximally
entangled basis on $d\otimes d$, where
$$\Phi_{m,n}={1}/{\sqrt{d}}\sum_{k=0}^{d-1}\omega^{kn}\ket{k}\ket{k+m~
{\rm mod} ~d},~\omega=e^{\frac{2\pi i}{d}}.$$ Let
$S=\{\Phi_{m,n}:mn=0\}$.  We claim that $S$ is unambiguously
distinguishable by LOCC. This is due to the fact that $span(S)$ can
also be spanned by a minimal GUPB,  i.e.,  any $2d-1$ states of the
following set $\{\ket{m}\ket{m}:0\leq m\leq d-1\}\cup
\{\ket{\overline{m}}\ket{\overline{m}}:0\leq m\leq d-1\}$, where
$\ket{\overline{m}}= 1/\sqrt{d}\sum_{k=0}^{d-1}\omega^{km}\ket{k}$.
Here we have employed Example $1$ and Theorem $1$.

Let us now consider the following question: What kind of basis for
$\mathcal{H}$ can be unambiguously discriminated by LOCC?
Furthermore, can we find a locally DEB? Surprisingly, we shall show
there do exist a DEB. Indeed we have a more general result: An
equivalence between a DEB and an IPB. Then the existence of DEB
follows directly from the existence of IPB. To state this
equivalence, we need the notion of reciprocal basis. Suppose
$S=\{\Psi_1,\cdots,\Psi_d\}$ is a basis for $\mathcal{H}$. For each
$1\leq k\leq d$, we can uniquely determine the reciprocal state
$\widetilde{\Psi}_k$ of $\Psi_k$ as follows:
$\ip{\widetilde{\Psi}_k}{\Psi_j}=0$ for any $j\neq k$. Then the
reciprocal basis for $S$, denoted by $\widetilde{S}$, is just the
collection of $\widetilde{\Psi}_k$.

{Theorem 3:} Let $S$ be a basis for $\mathcal{H}$. Then $S$ is
unambiguously distinguishable by LOCC if and only if $\widetilde{S}$
is a product basis. Furthermore, $S$ is a DEB if and only if
$\widetilde{S}$ is an IPB.

{Proof:} This result is essentially due to Lemma 3 and the following
fact: The reciprocal basis of $\widetilde{S}$ is just $S$ (up to
some unimportant phase factors), i.e.,
$\widetilde{\widetilde{S}}=S$.\hfill $\square$

We have presented an analytical construction of IPB in the proof of
Theorem 2. By the above theorem, a DEB can be obtained by taking its
reciprocal basis.  An illustrative example of DEB on $2\otimes 2$ is
as follows:

{Example 2:} Consider the following four states:
$\Psi_1=\ket{0}\ket{0}, \Psi_2=\ket{1}\ket{1},
\Psi_3=\ket{+}\ket{+}, \Psi_4=\ket{i_+}\ket{i_-},$ where $\ket{
i_{\pm}}=1/\sqrt{2}(\ket{0}\pm i\ket{1})$.
$S=\{\Psi_1,\Psi_2,\Psi_3,\Psi_4\}$ is a product basis for $2\otimes
2$ and any proper subset of $S$ with three members is a GUPB. Thus
$S$ is an IPB. By Theorem 3, the reciprocal basis $\widetilde{S}$ is
a DEB. The members of $\widetilde{S}$ are calculated as follows:
$$\widetilde{\Psi}_1={1}/{\sqrt{2}}\ket{00}-{1}/{2}\Phi_{+}+{1}/{2}i\Phi_{-},$$
$$\widetilde{\Psi}_2={1}/{\sqrt{2}}\ket{11}-{1}/{2}\Phi_{+}+{1}/{2}i\Phi_{-},$$
and $\widetilde{\Psi}_{3,4}=\Phi_{\pm}={1}/{\sqrt{2}}\ket{01}\pm
{1}/{\sqrt{2}}\ket{10}.$ These states are unambiguously
distinguishable by LOCC as for each $1\leq k\leq 4$, $\Psi_k$ is
just a product detecting state for $\widetilde{\Psi}_k$.

In conclusion, we introduce the notion of unextendible bases and
employ it to study the local distinguishability of multipartite
quantum states. In particular, a tight lower bound  on the number of
locally unambiguously distinguishable members of an arbitrary basis
is presented. We also obtain an equivalence between DEB and IPB, and
exhibit analytical constructions of such special basis. This
equivalence motivates  us to consider the relation among
orthogonality, entanglement, and local distinguishability. Roughly
speaking, neither orthogonality nor entanglement can uniquely
determine the distinguishablity of a basis. There may exist a
tradeoff: More orthogonality and less entanglement (or vice versa)
would sometimes enhance the local distinguishability of quantum
states. A challenging problem left is to obtain similar bounds in
the context of exact LOCC discrimination. (Partial results have been
obtained in Ref. \cite{HMM+06}).

We thank J.-X. Chen, G.-M. Wang, Z.-H. Wei, and C. Zhang for helpful
discussions. We also thank Dr. S. Bandyopadhyay for pointing out a
typographical error in the previous version of this Letter. This
work was partly supported by the Natural Science Foundation of China
(Grant Nos. 60621062 and 60503001), the Hi-Tech Research and
Development Program of China (863 project) (Grant No. 2006AA01Z102),
and the National Basic Research Program of China (Grant No.
2007CB807901).

\end{document}